\documentstyle[preprint,tighten,aps,amssymb,epsfig]{revtex}

\def\SU{ {SU}}
\def\su{{su}} 
 
\def\so{ {so}}

\def\R{{\Bbb R}}

\def\g#1{{\scriptscriptstyle (\!  #1 \!  )}}
\def\<{\langle}
\def\>{\rangle}
\newtheorem{theo}{Theorem}

\begin{document}

\preprint{
\vbox{
\hbox{\small Preprint CQG-98/05-02}
\hbox{\small Parma U. UPRF-98-04}
}
}

\title{$so(4)$ Plebanski Action and Relativistic Spin Foam Model}

\author{
R.\ De Pietri$^{1,2}$ and L.\ Freidel$^{2,3}$ \\[4mm]
$^{1}$ Dipartimento di Fisica, Universit\`a di Parma and 
INFN, Gruppo Collegato di Parma,\\
Viale delle Scienze, I-43100 Parma, Italy. \\
depietri@pr.infn.it \\[2mm]
$^{2}$ Center For Gravitational Physics and Geometry,
Physics Department, \\
The Pennsylvania State University,
104 Davey Laboratory, University Park, PA 16802 USA.\\
freidel@phys.psu.edu \\[2mm]
$^{3}$ Laboratoire de Physique Th\'eorique ENSLAPP 
Ecole Normale Sup\'erieure de Lyon,\\
46, All\'ee d'Italie, 69364 Lyon Cedex 07, France. \\
freidel@enslapp.ens-lyon.fr \\[2mm]
}

\maketitle

\begin{abstract}
In this note we study the correspondence between the ``Relativistic
Spin Foam'' model introduced by Barrett, Crane and Baez and the
$so(4)$ Plebanski action.  We argue that the $so(4)$ Plebanski model
is the continuum analog of the relativistic spin foam model.  We prove
that the Plebanski action possess four phases, one of which is gravity
and outline the discrepancy between this model and the model of
Euclidean gravity.  We also show that the Plebanski model possess
another natural discretisation and can be associated with another, new,
spin foam model that appear to be the $so(4)$ counterpart of the spin
foam model describing the self dual formulation of gravity.
\end{abstract}
\vskip.5pc
\centerline{pacs: 04.60.-m,04.60.Gw}
\vskip2pc



\section{Introduction}
 
The relativistic spin foam model was introduced by Barrett, Crane
\cite{BC97} and Baez \cite{B97} as a proposal for a state
sum formulation of 4-D Euclidean quantum gravity.  
These authors have analyzed the
classical description of a 4-simplex in terms of bivectors, quantized
this space and deduced from it a state sum model (or spin foam model
in the terminology of Baez).  Despite its deep geometrical meaning,
this model was not (precisely) interpreted as the partition function
associated with a definite classical action.
 
In this paper we show that the continuum action corresponding to the
relativistic spin foam model \cite{BC97,B97}, is given by the $so(4)$
Plebanski \cite{P77} action. The $so(4)$ Plebanski model was
originally introduced for the study of the possible relation between
the various Half-Flat (self-dual) solutions of complexified General
Relativity.  The action of this model is given by a constrained
$so(4)$ $BF$ model and its use in the context of classical and quantum
gravity has a long history \cite{Altri}. 

In section {\ref{BBC}} we recall some of the main results of
\cite{BC97,B97}. Namely the description of the 4-simplex in terms of
bivectors satisfying some constraints, the space of solutions of these
constraints and their quantization.  We then recall some facts
concerning $BF$ theory, state sum models, and present the relativistic
state sum in a form suitable for further analysis.  In section
{\ref{PA}} we present and study the $so(4)$ Plebanski model.  We show
that, when the $B$ field is non degenerate, this model consists of
four different sectors, one of which is gravity.  Performing a short
analysis of the quantization of the $so(4)$ Plebanski action we show
how these sectors interfere and emphasize the resulting discrepancy
between this model and Euclidean quantum gravity.  In section
{\ref{DC}} we study the discretization of the Plebanski action and the
corresponding state sum models.  We show that there exists two
possible discretizations of the model which lead to two different
state sum models.  One state sum model is the Baez-Crane-Barrett
model, while the other spin foam model is the $so(4)$ analog of the
Reisenberger spin foam model \cite{R97} (which corresponds to
self-dual formulation of gravity).  We inform the reader that after
this work was finished we becomed aware that some of these results
have been independently obtained by Reisenberger \cite{Rei}.


\section{The Baez-Barrett-Crane model}
\label{BBC}
In this section we recall the work done by Barrett and Crane
\cite{BC97} and further developed by Baez \cite{B97}.  We refer the
reader to their works for a deeper understanding of the model.  We
outline here their construction for reader's convenience.

We consider, as in \cite{BC97,B97}, the description of the geometry of
a 4-simplex in terms of bivectors $B$ associated to each 2-simplex
(triangle).  The properties of the bivectors for a non degenerate
4-simplex are
\begin{description}
\item[1] The bivector changes sign if the orientation of the triangle is
      changed.
\item[2] Each bivector satisfy the  simplicity condition i.e.
     \begin{equation}
         B\wedge B =0
     \end{equation}
\item[3] If two triangles share a common edge, then the sum of the two
     bivectors also satisfy the simplicity condition.
\item[4] The sum of the 4 bivectors corresponding to the faces of a
      tetrahedron is zero.  This sum is calculated using the
      orientation of the triangles given by the boundary of the
      tetrahedron.
\item[5] The assignment of bivectors to faces is non-degenerate.  This
      means that for six triangles sharing a common vertex, the six
      bivectors are linearly independent.
\end{description}

Each geometric 4-simplex determines a set of bivectors satisfying 
these conditions by defining $B=e\wedge e$, where $e$ are vectors 
associated with the edges of the 4-simplex.  Conversely,  each set of 
bivectors satisfying these conditions admits four types of solutions~: \\
${\rm I}^{\pm}$~: $B= \pm e\wedge e$, \\
${\rm II}^{\pm}$~: $B= \pm *(e\wedge e)$. \\
These bivectors can be identified with elements of the Lie algebra
$\so(4)$. (Also, it may be profitable to think of them as elements
of the dual of this Lie algebra.)  The splitting $\so(4)\simeq
\su(2)\oplus\su(2)$ is then the same as the splitting of the space of
bivectors into self-dual and anti-self-dual parts,
$\R^4\wedge\R^4=\Lambda^2_+\oplus\Lambda^2_-$. The condition that a
bivector $B$ is simple is
$$0=\<B,*B\>=\<B^+,B^+\>-\<B^-,B^-\>,$$ so that the norm of the self-dual 
and anti-self-dual parts is equal.  The area $A$ of the triangle is 
given by
$$A^2=\<B,B\>=\<B^+,B^+\>+\<B^-,B^-\>.$$

The quantization of the model is obtained by labeling 
the triangles with irreducible representations of 
$\SU(2)\times\SU(2)$, and each tetrahedron is labelled with a tensor 
in the product of the four spaces on its faces.
Irreducible representations of $\SU(2)\times\SU(2)$
 are pairs $(j,k)$ of representations of $\SU(2)$.  
The quantum analog of the properties 1--4 of bivectors in a 4-simplex 
are given by~:
\begin{description}
\item[Q1] Changing the orientation of a triangle changes the
      representation to its dual. 
\item[Q2] The representations on the triangles are of the form
      $(j,j)$. These are called simple representations.
\item[Q3] For any pair of faces of a tetrahedron, the pair of
      representations can be decomposed into its Clebsch-Gordan series
      for $\SU(2)\times\SU(2)$. Under this isomorphism, the tensor for
      the tetrahedron decomposes into summands which are non-zero only
      for the simple representations of $\SU(2)\times\SU(2)$.
\item[Q4] The tensor for the tetrahedron is an invariant tensor.
\end{description}

Baez, Barrett and Crane \cite{BC97,B97} constructed a state sum model
incorporating the previous quantum constraints.  Let us recall first
that there is a state sum formulation of the topological $BF$ theory
in 4-dimensions with (or without cosmological constant) given by the
Crane-Yetter \cite{CY93} or Ooguri \cite{O92} models.  The partition
function of the model associated to a triangulation $\Delta$ of a
manifold $M$ can be written as~:
\begin{equation}
Z_{BF}(\Delta, \Lambda) = \sum_{j_f,i_t} \prod_{f} dim_q(j_f) 
\prod_{v} \phi_{q,v}(\vec \jmath, \vec \imath),
\end{equation}
where $q=\exp ({\rm i}\Lambda)$, $j_f$ denotes a coloring of the faces
of $\Delta$ by irreducible representation of $U_q(su(2))$, $i_t$
denotes a coloring of the tetrahedra of $\Delta$ by intertwiners and
the sum is over all such colorings with $j\leq{ 2\pi \over
\Lambda}$. Moreover $dim_q(j)$ denotes the quantum dimension of the
representation of spin $j$ and $\phi_v(\vec \jmath, \vec \imath)$
denote the quantum $15$-j symbol associated with the 4-simplex $v$.
More precisely, associated to a 4-simplex $v$ we consider a graph
$\Gamma_v$ given by the intersection between the 2-skeleton of the
complex which is dual to the triangulation and the boundary of the
4-simplex $v$.  $\Gamma_v$ corresponds to the pentagram graph and we
color its 10 edges by $\vec j$ and its 5 vertices by $\vec i$.
$\phi_{q,v}(\vec \jmath, \vec \imath)$ corresponds to the
Reshetikhin-Turaev evaluation of the colored graph $\Gamma_v$.  This
state sum does not depend on the triangulation and corresponds to the
evaluation of the partition fuction
\begin{equation}
Z(M, \Lambda)=\int\!\! {\cal D}A {\cal D}B ~
  e^{\textstyle i\int\!\!_M tr (B \wedge F(A)-{\Lambda\over 2} B\wedge B)},
\label{BB}
\end{equation}
where $F(A)$ is the curvature of the $su(2)$ connection $A$.
 
The Baez, Barrett and Crane model is a modification of the state sum
(\ref{BB}) in the case of the $so(4)$ gauge group in which the
conditions corresponding to the quantum 4-simplex conditions (when
$\Lambda=0$) are imposed by hand. Yetter generalized this construction
to the case of non vanishing $\Lambda$ \cite{Yet}.  Yetter emphasized
that the quantum group that should be considered is $U_q(su(2))
\otimes U_{q^{-1}}(su(2))$.  This state sum model can be written~:
\begin{eqnarray}\label{BC}
Z_{BC}(\Delta, \Lambda) 
  &=& \sum_{j_f,\jmath^\prime_f,i_t, \imath^\prime_t } \prod_{f} 
      dim_q(j_f) dim_{q^{-1}}(\jmath^\prime_f) 
\prod_{v}  \phi_{q,v}(\vec \jmath, \vec \imath) 
      \phi_{q^{-1},v}(\vec{\jmath^{\prime}}, {\vec {\imath^\prime}}) 
      \delta_{\vec \jmath, \vec{\jmath^{\prime}}}\delta_{\vec \imath, 
      \vec{\imath^\prime }}.  
\end{eqnarray}
So this model corresponds to two copies of $su(2)$ $BF$ theories (or
an $so(4)$ $BF$ theory) together with $15$ constraints imposed for
each 4-simplex.  The presence of $\delta_{\vec j, \vec {\jmath^\prime
}}$ corresponds to the constraint {\bf Q2} while $\delta_{\vec i,
\vec{\imath^\prime }}$ to the constraint {\bf Q3}.  We will give the
continuum version of this partition function in equation (\ref{part}).

\section{The $so(4)$ Plebanski action}
\label{PA}
We use the notation given in the appendix.  The $so(4)$ Plebanski
action \cite{P77}, which depends on an $so(4)$ connection $\omega=
\omega_\mu^{IJ}X_{IJ} dx^\mu$, a two form valued into $so(4)$ $B=
B^{IJ}_{\mu\nu} X_{IJ} dx^\mu\wedge dx^\nu$, and a scalar symmetric
traceless matrix $\phi_{[IJ][KL]}$, ($\epsilon^{IJKL} \phi_{IJKL}=0$),
reads:
\begin{eqnarray}
  {\cal S}[\omega;B;\phi] &=& \int_M\!\!\!  
  \bigg[ B^{IJ}\wedge F_{IJ}(\omega) 
- \frac{\Lambda}{4}\epsilon_{IJKL} B^{IJ} \wedge B^{KL}
- \frac{1}{2} \phi(B)_{IJ}\wedge B^{KL}\bigg] ~,
\label{AP}
\end{eqnarray}
where $\phi(B)_{IJ}= \phi_{IJKL} B^{KL}$.  The whole set of 
Euler-Lagrange equations associated to the Plebanski action is given 
by:
\begin{eqnarray}
\frac{\delta {\cal S}}{\delta \omega^{\g{\alpha}}_\mu}
   &\rightarrow&
  DB=dB+[\omega,B] = 0, \\
\frac{\delta {\cal S}}{\delta B^{\g{\alpha}\g{\beta}}_{\mu\nu}}
   &\rightarrow& 
   F^{IJ}[ \omega ] = \frac{\Lambda}{2} {\epsilon^{IJ}}_{KL} B^{KL} 
                     + \phi(B)^{IJ}, \\
\frac{\delta {\cal S}}{\delta \phi^{\g{\alpha}\g{\beta}}}
   &\rightarrow& 
    B^{IJ}\wedge B^{KL} = {e}\epsilon^{IJKL} ~~,
    \label{ELsix}
\end{eqnarray}
where
\begin{equation}
  e = \frac{1}{4!} \epsilon_{IJKL} B^{IJ}\wedge B^{KL} ~~.
\end{equation} 
We can rewrite the action using the decomposition of the fields into
self-dual and anti self-dual fields, using the the duality in the Lie
algebra~: $B= B^\g{+} + B^{\g{-}}$, $\omega= \omega^\g{+} +
\omega^{\g{-}}$,and $\phi =\phi^\g{+}+ \psi+ \phi^\g{-}+\phi_{0}$
(see eq. (\ref{Aphi}) in appendix).
Here we have decomposed the Lagrange-multiplier field
$\phi_{IJKL}$ into its irreducible components with 
respect to the symmetry group: $\phi^\g{+}_{ij}$ (2,0) 5 components 
(left part of the Weyl); $\psi_{ij}$ (1,1) 9 components (traceless 
part of the Ricci); $\phi^\g{-}_{ij}$ (0,2) 5 components (right part of 
the Weyl); $\phi_{0}$ (0,0) 1 component (scalar curvature).  The 
action (\ref{AP}) decomposes into three parts.
\begin{eqnarray}
  {\cal S}           &=& {\cal S}^{+} + {\cal S}^{0} + {\cal S}^{-}  \\
  {\cal S}^{\pm} &=& \int_{M}\!\!  \bigg[  \delta_{ij} B^{\g{\pm}i}\wedge 
  F^{\g{\pm}j} - \frac{\phi_0 \pm \Lambda }{2}  \delta_{ij} B^{\g{\pm}i}\wedge 
  B^{\g{\pm}j} 
-\frac{1}{2} \phi^\g{\pm}_{ij} B^{\g{\pm}i} \wedge 
      B^{\g{\pm}j} \bigg] 
\\
  {\cal S}^{0} &=& \int_M \!  
      \bigg[- \psi_{ij} B^{\g{-}i}\wedge B^{\g{+}j} \bigg]
\end{eqnarray} 
Note that the Ashtekar formulation is derived by considering only the 
self-dual (left) part of the Plebanski action ${\cal S}^{+}$ or by 
imposing $B^{\g{+}}=0$.

The Euler-Lagrange equations \ref{ELsix} can be rewritten as:
\begin{eqnarray}
\frac{\delta {\cal S}}{\delta \phi^\g{+}_{ij}}
   &\propto& B^{\g{+}i}\wedge B^{\g{+}j} -\frac{1}{3} \delta^{ij} \delta_{kl} 
       B^{\g{+}k}\wedge B^{\g{+}l}  = 0 \label{ELwL} \\
\frac{\delta {\cal S}}{\delta \phi^\g{-}_{ij}}
   &\propto& B^{\g{-}i} B^{\g{-}j} -\frac{1}{3} \delta^{ij} \delta_{kl} 
       B^{\g{-}k}\wedge B^{\g{-}l} = 0 \label{ELwR} \\
\frac{\delta {\cal S}}{\delta \psi_{ij}} &\propto& B^{\g{-}i}\wedge 
B^{\g{+}j} = 0 \label{ELricci} \\
\frac{\delta {\cal S}}{\delta \phi_0}
   &\propto&  \delta_{ij}  \left[
       B^{\g{+}i}\wedge B^{\g{+}j} +
       B^{\g{-}i}\wedge B^{\g{-}j} \right] = 0 ~~. \label{ELtr} 
\end{eqnarray}
The set of equations (\ref{ELwL},\ref{ELwR},\ref{ELricci},\ref{ELtr})
is of course equivalent to equation (\ref{ELsix}), with
\[
\frac{1}{3} \delta_{kl} B^{\g{+}k}\wedge B^{\g{+}l} =
-\frac{1}{3} \delta_{kl} B^{\g{-}k}\wedge B^{\g{-}l} = 2 e
.
\]
We can now state the following:
\begin{theo}\label{dec} 
If
\begin{equation}
  \tilde{e} = \frac{1}{4!} \epsilon_{IJKL} \epsilon^{\mu\nu\rho\sigma} 
  B^{IJ}_{\mu\nu} B^{KL}_{\rho\sigma} \neq 0 ~~,
\end{equation}
then equation (\ref{ELsix}) is equivalent to equation
\begin{equation}
\epsilon_{IJKL} B^{IJ}_{\mu\nu} B^{KL}_{\rho\sigma} = 
\epsilon_{\mu\nu\rho\sigma} \tilde{e}  ~.
 \label{ELsixbis}
\end{equation}
Moreover, in this case, (\ref{ELsix}) and (\ref{ELsixbis}) are
fullfilled iff there  exists a real tetrad field 
$e^I= e^I_\mu dx^\mu$ such that one among the 
following equalities is satisfied~:
\begin{eqnarray*}
{\rm I}^+  &\qquad & B^{IJ} = +e^{I}\wedge e^{J} \\
{\rm I}^-  & &B^{IJ}= -e^{I}\wedge e^{J} \\
{\rm II}^+ & &B^{IJ}= +\frac{1}{2} {\epsilon^{IJ}}_{KL}e^{I}\wedge e^{J} \\
{\rm II}^- & &B^{IJ}= -\frac{1}{2} {\epsilon^{IJ}}_{KL}e^{I}\wedge e^{J} 
\end{eqnarray*}
\end{theo}
{\bf Proof}\\
The proof of the equivalence of conditions (\ref{ELsix}) and (\ref{ELsixbis})
when $\tilde{e}\neq 0$ is quite simple. In fact, we can define:
\begin{equation}
  \Sigma^{\mu\nu}_{IJ} = \frac{1}{\tilde{e}} \epsilon^{\mu\nu\rho\sigma} 
  \epsilon_{IJKL} B^{KL}_{\rho\sigma}  ~~,
\end{equation}
and in terms of $B$ and $\Sigma$ conditions (\ref{ELsix}) and 
(\ref{ELsixbis})
 respectively read:
\begin{eqnarray}
  \Sigma^{\mu\nu}_{IJ} B^{IJ}_{\rho\sigma} = \delta^{\mu\nu}_{\rho\sigma} 
  \\
  \Sigma^{\mu\nu}_{IJ} B^{KL}_{\mu\nu} = \delta^{KL}_{IJ}.
\end{eqnarray}
It is easy to verify the {\it if} part of the theorem, and to see that 
equation (\ref{ELsix}) is invariant under the change $B \rightarrow
-B$ or $B\rightarrow *B$.  The {\it only if} part can be proved as follows.
From Reisenberger \cite{R95}, we know that equation (\ref{ELwL}) (resp 
eq.  (\ref{ELwR})) implies that there exists a real tetrad field 
$a^I_\mu$ (resp.  $b^I_\mu$) such that $B^{\g{+}i} =\pm 
T^{\g{+}i}_{IJ} a^I\wedge a^J$ (resp.  $B^{\g{-}i} =\pm 
T^{\g{-}i}_{IJ} b^I\wedge b^J$).  The condition $\tilde{e}\neq 0$ implies that 
$a^I$ is a basis of one-forms, so there exists a $4\times 4$ real 
matrix $ A$ such that~: 
\begin{equation} b^I = A^I_J a^J.
\end{equation}
Equation (\ref{ELtr}) is equivalent to the requirement that
$det(A) =\pm 1$ so that $\pm A \in SL(4)$.  Equation (\ref{ELricci}) is
equivalent to the the requirement that $\pm A \in SO(4)$.  Using the
decomposition $SO(4)= (SU(2)\times SU(2))/{Z_2}$ we can decompose $A$ into a 
commuting product of self-dual and anti-self dual rotations, $\pm 
A=O_{+}^{-1} O_{-}$.  If we define $e^I = {O_{+}}^I_J b^J = \pm 
{O_{-}}^I_J a^J $, then
\begin{equation}
B^{\g{+}i} =\pm T^{\g{+}i}_{IJ} e^I\wedge e^J ,\,\,
B^{\g{-}i} =\pm T^{\g{-}i}_{IJ} e^I\wedge e^J.
\end{equation}
$\framebox[3mm]{}$ \\

It is interesting to note that given an $so(4)$ two-form field
$B^{IJ}$, it is possible to construct two Urbantke
\cite{U83} $su(2)$ metrics
\begin{eqnarray}
g_{\mu\nu}^\g{+} 
  &=& -\frac{2}{3\tilde{e}}
    \epsilon_{ijk} \epsilon^{\alpha\beta\gamma\delta}
    \left[ B^{\g{+}i}_{\mu\alpha} 
           B^{\g{+}j}_{\beta\gamma} 
           B^{\g{+}k}_{\delta\nu} \right] \\
{g}_{\mu\nu}^\g{-} 
 &=& -\frac{2}{3\tilde{e}}
     \epsilon_{ijk} \epsilon^{\alpha\beta\gamma\delta}
          \left[ B^{\g{-}i}_{\mu\alpha} B^{\g{-}j}_{\beta\gamma} 
          B^{\g{-}k}_{\delta\nu} \right]  ~~.
\end{eqnarray}
For a general field $B$ these two metrics are unrelated, but when 
equation (\ref{ELsix}) is satisfied they are equal up to a sign:
\begin{eqnarray*}
{\rm I}^+ & \qquad &
       ~g_{\mu\nu}^\g{+} 
   = - ~g_{\mu\nu}^\g{-} 
   = \eta_{IJ} e^I_{\mu} e^J_{\nu} \\
{\rm I}^- & & 
       ~g_{\mu\nu}^\g{+} 
   = - ~g_{\mu\nu}^\g{-}
   = - \eta_{IJ} e^I_{\mu} e^J_{\nu} \\
{\rm II}^+ & &
       ~g_{\mu\nu}^\g{+} 
     = ~g_{\mu\nu}^\g{-} 
     = \eta_{IJ} e^I_{\mu} e^J_{\nu} \\
{\rm II}^- & &
      ~g_{\mu\nu}^\g{+} 
     =~g_{\mu\nu}^\g{-}
     = - \eta_{IJ} e^I_{\mu} e^J_{\nu} 
\end{eqnarray*}

\noindent
Case I corresponds to the sector where the action becomes:
\begin{eqnarray}
{\cal S}_{{\rm I}^\pm}[\omega^{IJ}_\mu;e^I_{\mu}] &=& \int\!\! 
\bigg[\pm e^I \wedge e^J\wedge F_{IJ}[\omega] +
\frac{\Lambda}{4} \epsilon_{IJRS} e^I\wedge e^J\wedge e^R 
\wedge e^S \bigg], 
\end{eqnarray}
and case II to the gravitational sector~:
\begin{eqnarray}
 {\cal S}_{{\rm II}^\pm}[ \omega^{IJ}_\mu;e^I_{\mu}] &=& \int_M \bigg[ 
\pm  \epsilon_{IJRS} e^I\wedge e^J \wedge F^{RS}[\omega] + 
\frac{\Lambda}{4} \epsilon_{IJRS} e^I\wedge e^J\wedge e^R\wedge 
e^S \bigg].  
\end{eqnarray}


After integration over the field $\phi$,
the partition function of the Plebanski model reads:
\begin{eqnarray}
\label{part}
Z_{Pl}(M, \Lambda) 
&=& \int\!\! 
  {\cal D}A^\g{+} {\cal D}A^\g{-} {\cal D}B^\g{+}{\cal D}B^\g{-} 
  ~\delta(f(B))  \times \\
&&~~ e^{\textstyle i\int_M tr (B^\g{+} \wedge F(A^\g{+}) 
        -{\Lambda\over 2} B^\g{+}\wedge B^\g{+})}  
   \times \nonumber \\
&&~~ e^{\textstyle i\int_M tr (B^\g{-} \wedge F(A^\g{-}) 
        +{\Lambda\over 2} B^\g{-}\wedge B^\g{-})} ,
\nonumber
\end{eqnarray}
where $\delta(f(B))$ is a $20$ dimensional delta function 
corresponding to the set of constraints (\ref{ELsix}).
This expression clearly appears to be 
the continuous analogous of (\ref{BC})
(we will prove this relation in the next section). 
The fact that the $(+)$ and $(-)$ parts 
have opposite cosmological constant (which is the only possibility
due to the constraint (\ref{ELtr})) is the field-theoretic motivation for 
the Yetter \cite{Yet} choice of the quantum group $U_q(su(2)) \otimes 
U_{q^{-1}}(su(2))$.

Using the result of Theorem \ref{dec}, this partition 
function can be written (when $\Lambda=0$)
\begin{eqnarray}
&& Z_{Pl}(M, \Lambda) = \!\int\!\!\! {\cal D}A {\cal D}E
\bigg[ \cos ({\int_{M}\! e^I \wedge e^J\wedge F_{IJ}}) 
+ \cos ({\int_{M}\! \epsilon_{IJRS}e^I \wedge e^J\wedge F_{RS}})
\bigg].
\label{Ppf}
\end{eqnarray}
In this expression we have neglected contributions from degenerate $B$'s 
($e=0$) (which is not justified at this point). 
The expression (\ref{Ppf}) clearly shows the difference between the 
Plebanski model and pure gravity.  In the Plebanski model, even if we 
consider only non degenerate $B$, we still integrate over configurations of the 
$B$ fields that can globally belong to different sectors, which results 
in interference between different sectors. 
We have seen previously that the $so(4)$ Plebanski model possess a
$Z_2 \times Z_2$ symmetry $B\rightarrow -B$, $B\rightarrow *B$, this 
discrete symmetry exchange the different sectors  and is responsible 
for the existence of interference. The interference kill for instance 
the imaginary 
part of the amplitude. Gravity is obtained by 
restricting the $B$ fields to be always in the $II^+$ sector.  For 
example, the partition function of Euclidian pure gravity without 
cosmological constant is:
\begin{equation}
Z_{GR}= \!\int\!\!\!  {\cal D}A {\cal D}E 
e^{i \int_{M}\! \epsilon_{IJRS}e^I \wedge e^J\wedge F_{RS}}
\end{equation}
If we allow in the partition function degenerate configurations, the
situation is even worse because we should then integrate between
configurations of the $B$ fields which can belong to different sectors
at different points. 

This means that in all cases, 
there is a discrepancy between the Barret-Crane model 
($so(4)$ Plebanski model) and gravity due to interference
between different sector.
It is important to note that this discrepancy between $BF$ formulation
of gravity and Einstein theory already appears in the case of
3-dimensional Euclidian gravity  \cite{KL1} and also in the self-dual formulation
of 4-dimensional Euclidean gravity.

\section{The discretization of the constraints}
\label{DC}
In order to link the partition function (\ref{part}) to the
Baez-Barrett-Crane model we have to discretize the set of constraints
(\ref{ELsix},\ref{ELsixbis}) along 4-simplices.  Using the $B$ field,
we can associate an element of $so(4)$ to each 2-dimensional surface
$S$ embedded in $M^4$ as follows:
\begin{eqnarray}
	\label{dis} 
     {B}^{IJ}[S] = \int_S B^{IJ}_{\mu\nu} dx^\mu \wedge dx^\nu ~~.
\end{eqnarray} 
Given a triangulation of our manifold $M^4$ by 4-simplices, we take
the $B$ field to be constant inside each 4-simplex, so that $d
B^{IJ}=0$.  Using (\ref{dis}), we can associate a bivector to each
face (2-simplex) of the 4-simplex.  Then, by Stokes' theorem we see
that the sum of bivectors over all faces of a given tetrahedron
belonging to the 4-simplex is 0: this is the closure constraint {\bf Q4}.

We have seen, in theorem \ref{dec}, that in the continuum case the
constraints on the $B$ field can be written, when $B$ is non
degenerate, in two different but equivalent forms, (\ref{ELsix}) or
({\ref{ELsixbis}).  Namely
\begin{equation}
\epsilon_{IJKL}B^{IJ}_{\mu\nu} B^{KL}_{\rho\sigma}
= \tilde{e} \epsilon_{\mu \nu\rho\sigma},
\label{i}
\end{equation}
or
\begin{equation}
\epsilon^{\mu \nu\rho\sigma} B^{IJ}_{\mu\nu} B^{KL}_{\rho\sigma} = \tilde{e} 
\epsilon^{IJKL}.
\label{ii}
\end{equation}
It is important to understand that these two equivalent form of the
constraints lead to two different discretizations and then to two, a
priori different, state sum models.  As we will see, one state sum
model (or { spin foam model} in the terminology of Baez),
corresponding to (\ref{i}) is the Baez-Crane-Barrett model, while the
other spin foam model, corresponding to (\ref{ii}), is the $so(4)$
analog of the Reisenberger spin foam model \cite{R97} (which
corresponds to self-dual formulation of gravity).

Let us consider first the discretization of (\ref{i}).  The free
indices of (\ref{i}) are space-time indices so that we can saturate
them by integrating in all possible ways over faces of the
4-simplex. A straightforward computation show that (\ref{i}) implies:
\begin{eqnarray}
&&V(S,\tilde{S}) =  \epsilon_{IJKL} {B}^{IJ}[S] {B}^{KL}[\tilde{S}] 
= \delta_{ij} \left[ {B}^{\g{+}i}[S] {B}^{\g{+}j}[\tilde{S}] - 
   {B}^{\g{-}i}[S] {B}^{\g{-}j}[\tilde{S}] \right],
\nonumber \end{eqnarray}
where $V(S,\tilde{S})= \int_{x\in S, y\in \tilde{S}} \tilde{e}
\epsilon_{\mu \nu \rho \sigma} dx^\mu \wedge dx^\nu \wedge dy^\rho
\wedge dy^\sigma $ is the 4-volume spanned by $S$ and $\tilde{S}$.  
In particular, this means
\begin{equation}
   \delta_{ij} \left[ {B}^{\g{+}i}[S] {B}^{\g{+}j}[{S}] 
   - {B}^{\g{-}i}[S] {B}^{\g{-}j}[{S}] \right] =0,
\label{AA}
\end{equation}
for each face (2-simplex) $S$ of the 4-simplex (conditions Q2), and 
\begin{equation}
   \delta_{ij} \left[ {B}^{\g{+}i}[\tilde{S}] {B}^{\g{+}j}[{S}] 
   - {B}^{\g{-}i}[\tilde{S}] {B}^{\g{-}j}[{S}] \right] =0,
\end{equation}
for each couple of faces $S, \tilde{S}$ which share one edge (that,
combined with (\ref{AA}), impose condition Q3).  Therefore the
bivectors associated via (\ref{dis}) to the faces of the tetrahedra
satisfy the Baez-Barrett-Crane constraints of section \ref{BBC}.  We
can thus conclude that the state sum model (\ref{BC}) corresponds to a
natural discretization of the main sector of the $so(4)$ Plebanski
model.

We now consider the continuum constraint in the form (\ref{ii}).  In
this form of the constraint, the space-time indices are saturated by
the $\epsilon$ tensor while the internal indices are free.  A general
procedure is presented in \cite{KL} that describes the construction of
{ spin foam models} from classical actions of certain $BF$ type.  The
form (\ref{ii}) of the constraints allows us to use the results of
\cite{KL} from which we can deduce the discretization of (\ref{ii})
and the corresponding { spin foam model}.  For the reader's
convenience we nevertheless give a self contained justification.  The
discretization of the $B$ field inside a 4-simplex amounts to
decompose the 2-form $B$, inside the 4-simplex, into a sum of singular
2-forms associated with the faces of the 4-simplex~:
\begin{equation}
    B^{IJ}(x) =\sum_{S} B^{IJ}_{S}(x) ~~,
\end{equation}
where the sum is over all faces of the 4-simplex and $B^{IJ}_S$ is a 
two form such that:
\begin{equation}
     \int B^{IJ}_S \wedge J ={B}^{IJ}[S] \int_{S^*} J,
\end{equation}
where $J$ is any 2-form, and $S^*$ denote the dual face of $S$, i.e.,
the 2-cell dual to $S$ that belongs to the dual complex of the
4-simplex.  With this definition it is clear that~:
\begin{eqnarray}
&&\int_{S}B^{IJ}_{\tilde S} = \delta_{S, \tilde{S}} {B}^{IJ}[S], \\
&&\int_{M^4} B^{IJ}_{\tilde S}\wedge B^{KL}_{ S} = 
{B}^{IJ}[S] {B}^{KL}[\tilde{S}] \epsilon({S,\tilde{S}}),
\end{eqnarray}
where $\epsilon({S,\tilde{S}})$ is equal to the sign of the oriented
volume $V(S,\tilde{S})$ spanned by the 2 faces $S$ and $\tilde{S}$.
So $\epsilon({S,\tilde{S}})= \pm 1$ if $S, \tilde{S}$ are 2 faces of
the 4-simplex which do not share an edge and $\epsilon({S,\tilde{S}})= 0$
if $S, \tilde{S}$ do share an edge.  Using these definitions and
properties it is straightforward to see that the constraint (\ref{ii}),
after integration over the 4-simplex, leads to~:
\begin{equation}
\label{con}
\Omega^{IJKL} -\epsilon^{{IJKL}}
\frac{1}{4!}\epsilon_{ABCD}\Omega^{ABCD}=0  ~,
\end{equation}
where 
\begin{equation}
  \Omega^{IJKL} =\sum_{S, \tilde{S}} 
  \epsilon({S,\tilde{S}}) {B}^{IJ}[\tilde{S}] {B}^{KL}[{S}].
\end{equation}
This form of the constraint is analogous to the Reisenberger
constraint appearing as a discretization of the self-dual formulation of
gravity \cite{R96}.  And we can construct (as in \cite{R97}) the 
corresponding spin foam model (see also \cite{KL}).

If we decompose $\Omega$ into its self-dual and antiself-dual 
components the constraints ({\ref{con}}) are equivalent to~:
\begin{eqnarray}
\Omega^{ij}_{++} = \delta^{ij} \frac{1}{3} tr(\Omega_{++}) \label{c++}\\
\Omega^{ij}_{--} = \delta^{ij} \frac{1}{3} tr(\Omega_{--})  \\
\Omega^{ij}_{+-} = 0 \\
tr(\Omega_{++}) + tr(\Omega_{--}) =0 ~~,
\end{eqnarray}
where 
$ \Omega^{ij}_{\epsilon_1 \epsilon_2} = 
T^{\g{\epsilon_1}i}_{IJ}T^{\g{\epsilon_2}i}_{KL}\Omega^{IJKL}$.
These constraints are clearly the discrete analogous of constraints
(\ref{ELwL},\ref{ELwR},\ref{ELricci},\ref{ELtr}).
Moreover, (\ref{c++}) is precisely the Reisenberger constraint that 
appears in the { spin foam model} corresponding to self-dual formulation of 
gravity \cite{R96,R97}.

\section{Conclusion}

In this note, we have analyzed the connection between the
Barrett-Crane model and $so(4)$ Plebanski's action for Euclidean
General Relativity.  In particular, we have shown that the
Barrett-Crane conditions on the allowed $so(4)$ representation are the
quantum transcription of the Plebanski constraints (\ref{ELsixbis}).
This result implies that the Barrett-Crane state sum model is
associated to the partition function (\ref{Ppf}).  This model is
related to, but different from, pure gravity due to the presence of
interference terms between different sectors.  We also showed that the
$so(4)$ Plebanski model admits another possible dicretisation that is
analogous to the one already used for the self-dual formulation of
gravity.  Thus, it seems important to understand the link between the
Barrett-Crane constraint and the constraint (\ref{con}).  One would
expect that, at the quantum level, the space of solutions of the two
constraints are in one-to-one correspondence or that the two
corresponding state sum models converge to each other when the
triangulation becomes finer (if degenerate solutions are not
relevant).  We leave these questions for future analysis.
 
\acknowledgments

We thank Abhay Ashtekar, John Baez, Kirill Krasnov, Luca Lusanna,
Carlo Rovelli and Jose Zapata for many important suggestions and
valuable criticisms and insights.  This work has been partially
supported by the CNRS (France), by a NATO grant and by the INFN grant
``Iniziativa specifica FI-41'' (Italy).

\appendix
\section{Notation for the SO(4) and SO(3,1) groups.}

We use capital latin letter for internal vector index $I,J=0,1,2,3$
and the internal metric $\eta_{IJ}={\rm diag}(1,1,1,1)$.  
We denote with $\g{\alpha}$ the group indices and use as fundamental
basis
\begin{equation}
  T_\g{i}^{IJ}   = - \epsilon^{0iIJ} \qquad
  T_\g{i+3}^{IJ} = \eta^{iI} \eta^{0J} - \eta^{0I} \eta^{iJ}  
\end{equation}
In these two group the duality transformations is given by:
\begin{equation}
({}^*T_\g{\alpha})^{IJ} = \frac{1}{2} \epsilon^{IJ}_{~~KL} T_\g{\alpha}^{KL}
  \qquad
\begin{array}{l} {}^*T_\g{i}   = T_\g{i+3}  \\
                 {}^*T_\g{i+3} = T_\g{i} ~~.
\end{array}
\end{equation}
It is profitable to consider the duality basis
\begin{equation}
  T_{\g{+}i}^{IJ} = \frac{T_\g{i}^{IJ} + T_\g{i+3}^{IJ}}{2} \qquad 
  T_{\g{-}i}^{IJ} = \frac{T_\g{i}^{IJ} - T_\g{i+3}^{IJ}}{2}
\end{equation}
where we have
\begin{equation}
       {}^*T_{\g{+}i} = ~ T_{\g{+}i}  
\qquad {}^*T_{\g{-}i} = - T_{\g{-}i} 
\end{equation}
In this bases we have the relations
\begin{eqnarray}
  T_{\g{\pm}i}^{IK} T_{\g{\pm}j}{}_{K}^{~J} = 
    \frac{1}{2} \epsilon^{ijk} T_{\g{\pm}k}^{IJ} 
  - \frac{1}{4} \delta^{ij} \eta^{IK} \\
  \sum_{i} T_{\g{\pm}i}^{IJ} T_{\g{\pm}i}{}_{KL}
      = \frac{1}{2} \delta^{I}_{[K} \delta^{J}_{L]} \pm
        \frac{1}{4} \epsilon^{IJ}{}_{KL} 
\end{eqnarray}
and the duality projection of bivector is defined by:
\begin{equation}
B^{IJ} = T_{\g{+}i}^{IJ} B^{\g{+}i} + T_{\g{-}i}^{IJ} B^{\g{-}i}
~~.
\end{equation}
Finally, the decomposition of the Lagrange multiplier
field $\phi_{IJKL}$ is given by:
\begin{eqnarray}
\label{Aphi} 
  \phi_{IJKL} &=& 
     \phi^{\g{+}}_{ij} T^{\g{+}i}_{IJ} T^{\g{+}j}_{KL} 
   + \phi^{\g{-}}_{ij} T^{\g{-}i}_{IJ} T^{\g{-}j}_{KL} 
 \\ & & \nonumber
   + \psi_{ij} ( T^{\g{+}i}_{IJ} T^{\g{-}j}_{KL} 
                +T^{\g{-}j}_{IJ} T^{\g{+}i}_{KL}  )
\\ & & \nonumber
   + \phi_{0}~ \delta_{ij} ( T^{\g{+}i}_{IJ} T^{\g{+}j}_{KL} 
                +T^{\g{-}i}_{IJ} T^{\g{-}j}_{KL}  )
\end{eqnarray}
where $\phi^\g{+}_{ij}$ and $\phi^\g{-}_{ij}$ are symmetric and 
traceless. 



\begin{thebibliography}{10}

\bibitem{BC97} J.W.  Barrett and L.  Crane, {``Relativistic Spin 
Networks and Quantum Gravity''}, {\it J.  Math.  Phys.} {\bf 39}, (1998), 
3296.

\bibitem{B97} J.Baez, ``Spin Foam Models'',  
   {\it Class. quantum Grav.}, to appear.

\bibitem{P77} J.  F.  Pleba\`nski, { ``On the separation of Einstein 
Substructure''}, {\it J.  Math.  Phys.}  {\bf 12}, (1977), 2511.

\bibitem{Altri}
   R Capovilla, J Dell, T Jacobson, L Mason,  
      {\it Class. quantum Grav.} {\bf 8}, (1991), 41. \\
   R Capovilla, J Dell, T Jacobson,  
      {\it Class. quantum Grav.} {\bf 8}, (1991), 59. \\
   Obukhov Yu N. and Tertychin S. I., 
      {\it Class. Quantum Grav.} {\bf 13}, (1996), 1623. \\
   M. Pillin,
      {\it Class. quantum Grav.} {\bf 13}, (1996), 2329. 

\bibitem{R95}
   M. Reisenberger, {\it Nucl. Phys.} {\bf B457}, (1995), 643.

\bibitem{R97} 
   M. Reisenberger, ``A lattice worldsheet sum for 4-d Euclidean 
   general relativity'', {\it xxx-archive}: gr-qc/9711052.

\bibitem{Rei} 
   M.  Reisenberger, ``Classical Euclidean General Relativity from 
   'Lefthanded Area = Righthanded Area' '',
   {\it xxx-archive}: gr-qc/9804061.

\bibitem{CY93} L. Crane and D. Yetter, ``A categorical construction 
   of 4-d TQFTs'' in {\it Quantum Topology}  eds. L. Kauffman and
   R. Baadhio, World Scientific, Singapore, 1993.

\bibitem{O92} H.  Ooguri, ``Topological lattice model in four dimensions'' 
    {\it Mod.  Phys.  Lett} {\bf A7}, (1992), 2799.

\bibitem{Yet} D.Yetter, ``Generalised Barrett-Crane vertices 
   and invariants of embedded graphs'', {\it xxx-archive}: math.QA~/9801131.

\bibitem{U83} H. Urbantke, {\it J. Math. Phys.} {\bf 25}, (1983).
      

\bibitem{KL1} L.  Freidel and K.  Krasnov ``Discrete Space-Time Volume
       for 3-Dimensional BF Theory and Quantum Gravity'',
       {\it xxx-archive}: hep-th/9804184.

\bibitem{KL} L.  Freidel and K.  Krasnov ``Spin foam models and the 
classical action principle'', in redaction.

\bibitem{R96} 
   M.  Reisenberger, ``A left-handed simplicial action for Euclidean 
   general relativity'',{\it Class.  quantum Grav.} {\bf 14}, (1997), 1753.
   
\end{thebibliography}
\end{document}